\documentclass[nohyper,11pt,letterpaper]{JHEP3}
\usepackage[dvips]{epsfig}
\newcommand{\be} {\begin{equation}}
\newcommand{\ee} {\end{equation}}
\newcommand{\ba} {\begin{eqnarray}}
\newcommand{\ea} {\end{eqnarray}}
\newcommand{\n}[1]{\label{#1}}
\newcommand{\eq}[1]{Eq.(\ref{#1})}

\title{
Non-minimal scalar fields in 2D de Sitter and dilaton black holes
}

\author{
Andrei Zelnikov${}^{(a)}$\thanks{E-mail:
zelnikov@phys.ualberta.ca}\\
$^{(a)}$Theoretical Physics Institute, Department of Physics\\
University of Alberta, Edmonton, AB, Canada T6G 2J7
}

\date{\today}

\abstract{
We study non-minimal quantum fields in the gravitational field of
2-dimensional dilaton black holes and the de Sitter spacetime.
We found that the Green functions for non-minimal massless fields
in a particular class of dilaton black holes and in the de Sitter spacetime
are almost identical. Using this symmetry exact solutions are derived for
quasinormal modes and bound states in these background geometries.
The problem of stability of dilaton black holes is discussed.
}

\preprint{arXiv:0805.4031[hep-th]\\
 Alberta Thy 13-08}

\begin{document}

\section{Introduction}

Our purpose is to study quantum scalar fields living on the
background of a special class of solutions of (1+1)-dimensional
dilaton gravity. This class of  2D spacetimes includes a variety of
dilaton black holes (BH) such as CGHS solution \cite{CGHS} as well as
the de Sitter spacetime. In this paper non-minimality is understood
as a presence in the Lagrangian of an interaction term with the
scalar curvature rather than with the dilaton. We demonstrate that,
although the geometry of black holes is quite different from that of
the de Sitter space, the dynamics of quantum non-minimal massless
scalar fields does not depend really on this difference.
At first sight it looks surprising,
since the de Sitter geometry seems to be much more symmetrical than,
e.g., the CGHS model, but it happens that for non-minimal fields they
look alike. This property provides a tool to translate many
quantum effects, which were calculated for the de Sitter space,
to the 2D black hole background and
vice versa.  Quantum fields on the de Sitter background have been
studied in the literature in great detail and the Green functions for
scalar fields in the de Sitter space are well known. The only, though very
important, difference from the case of dilaton black holes is the
choice of boundary conditions for quantum states of the fields.
In the presence of interacting fields the de Sitter space (for
even dimensions) appears to be intrinsically unstable \cite{Polyakov:2007}.
Technically this is a consequence of an infrared asymptotic of the
Feynmann propagators corresponding to various vacua. But, as we prove
in the next section, the propagators on the de Sitter space are
identical to those on the considered class of dilaton black holes,
which strongly suggests their intrinsic instability too. Another
interesting property resulting from the discussed symmetry is that
quasi-normal modes for non-minimal fields on the background of 2D
dilaton black holes in question \eq{metric}
are identical to quasi-normal modes on the de Sitter space.

%%%%%%%%%%%%%%%%%%%%%%%%%%%%%%%%%%%%%%%%%%%%%%%%%%%%%%%%%%%%%%%%%%%%%%%%%%

\section{Dilaton models}\label{sec_1}
\setcounter{equation}0

Let us consider spacetimes described by the metric
\ba\n{metric}
ds^2=-4{dUdV\over(1-UV)^a}={{\mathrm e}^{x^*}
\over({\mathrm e}^{x^*}+1)^a}\left[-dt^2+d{x^*}^2\right].
\ea
where $U,V$ are the Kruskal coordinates, $x^*$ is the tortoise coordinate
and t is the time.
\ba\nonumber
U={\mathrm e}^{t+x^*\over 2}, \hskip 1cm V=-{\mathrm e}^{-{t-x^*\over 2}}, \hskip 1cm
x^*=\ln{(-UV)}.
\ea
For simplicity, we omit all dimensional prefactors in the metric,
so that the coordinates are dimensionless. Dimensional quantities
can be easily restored later by introducing a proper scale of the metric.

These spacetimes (\ref{metric}) naturally appear as generic static
solutions of a wide class of dilaton gravity models described by the action
\ba\n{dilaton}
S={1\over 16\pi}\int
d^2x\sqrt{-g}{\mathrm e}^{-2\phi}\left[R+4a(\nabla\phi)^2+B{\mathrm e}^{2(1-a)\phi}\right]
\ea
and parametrized by a dimensionless parameter $a$. Most of the physically
interesting spacetimes appear to belong to the interval $0\le a\le 2$.
The dimensional
parameter $B$ plays a role similar to the cosmological constant and
can be fixed by an appropriate rescaling of the metric. These dilaton
gravity models have been studied in detail by Fabbri  and Russo
\cite{FabbriRusso:96}. An excellent analysis of the problem of geodesic
completeness of these spacetimes and their generalizations can be
found in the paper by Katanaev, Kummer, and Liebl \cite{KatKumLie:96}.

Let us present a couple of other sets of static coordinates, which may be
more convenient for different applications
\ba\n{metric_1}
ds^2=-z(1-z)^{a-1}dt^2+{(1-z)^{a-3}\over z}dz^2
    =\left({1+r\over 2}\right)^{a-2}\left[-4(1-r^2)~dt^2+{dr^2\over 1-r^2}
\right],
\ea
\ba\nonumber
z=-{UV\over 1-UV}, \hskip 1cm x^*=\ln{z\over 1-z},
\hskip 1cm
r=1-2z.
\ea
For these metrics with an arbitrary parameter $a$ the horizon is located at $z=0$.
Surface gravity $\kappa$ on the horizon is
\ba\nonumber
\kappa=\sqrt{\Big|\xi^{\mu}\xi_{\mu}w^{\nu}w_{\nu}\Big|_{z=0}}={1\over
2}, \hskip 1cm \xi^{\mu}=\delta^{\mu}_t
\hskip 1cm w_\alpha={1\over
2}\nabla_{\alpha}\ln\left|\xi^{\mu}\xi_{\mu}\right|.
\ea
One would expect that quantum fields on this background reveal effective thermal
properties with the corresponding Hawking temperature to be related with the
horizon surface gravity
\ba\nonumber
T_{\mathrm BH}={\kappa\over 2\pi}={1\over 4\pi}.
\ea

\begin{figure}
\centerline{\includegraphics[width=6cm, height=4.5cm]{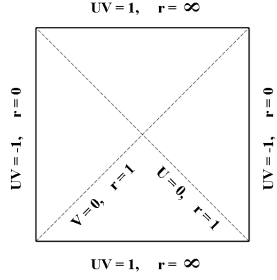}}
\caption{Carter-Penrose diagrams for the de Sitter spacetime
and geometries \eq{metric} with $1<a\le 2$}.
\label{fig1}
\end{figure}

\begin{figure}
\centerline{\includegraphics[width=6cm, height=4.5cm]{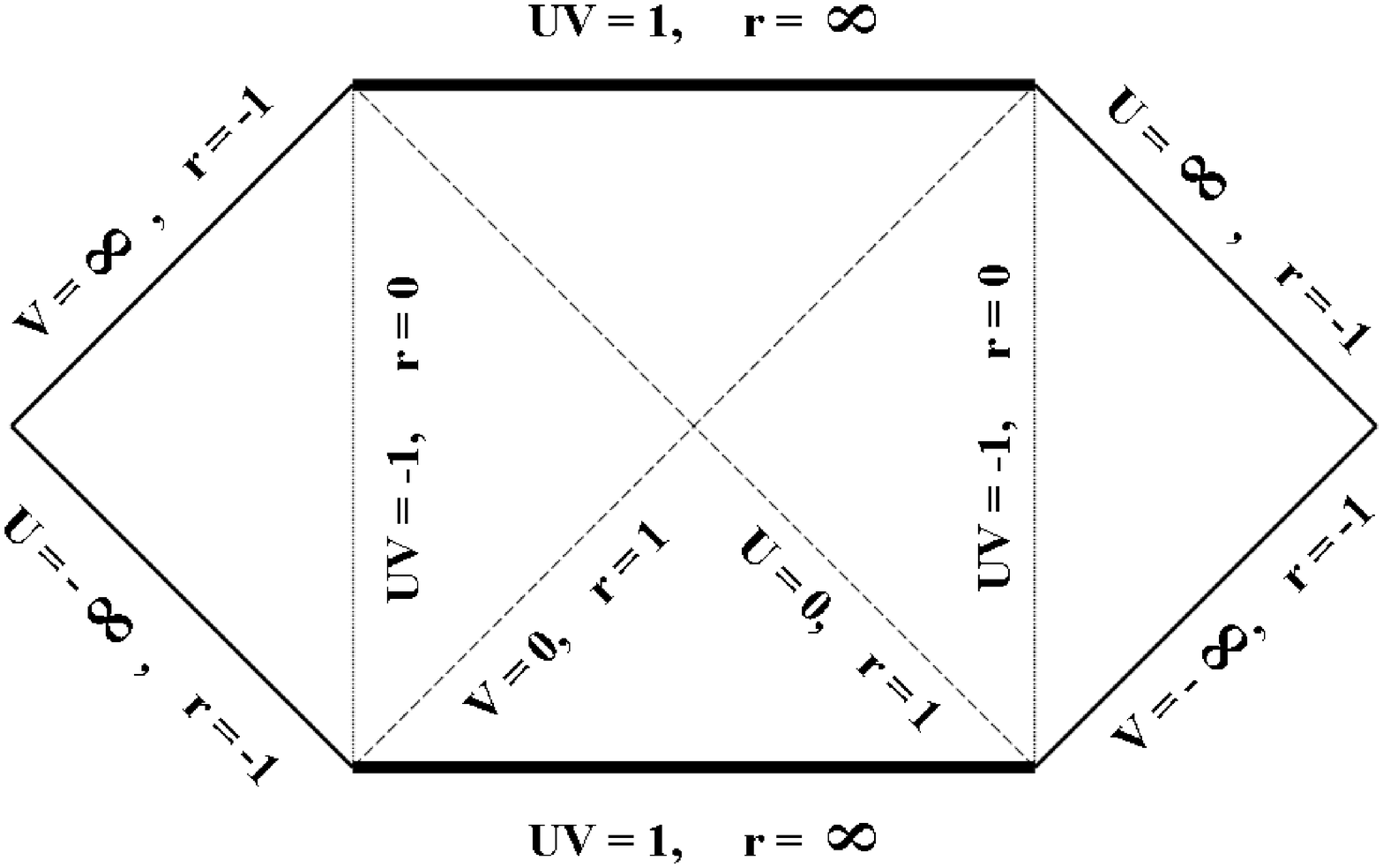}}
\caption{Carter-Penrose diagrams for the dilaton black holes with $0<a\le1$.}
\label{fig2}
\end{figure}

Carter-Penrose diagrams of these spacetimes are basically of two different types
\cite{KatKumLie:96} depending on the value of the parameter $a$.
If $1< a\le 2$ the global structure is similar to the de Sitter geometry
(see figure\ref{fig1}) while for
$0< a\le 1$ it is of the dilaton black hole type (see figure \ref{fig2}).
When the parameter $a=2$ the metric \eq{metric} describes the de Sitter spacetime.
\ba\nonumber
ds^2=-z(1-z)dt^2+{dz^2\over z(1-z)}
    =-(1-r^2)dT^2+{dr^2\over 1-r^2}, \hskip 1cm T={t\over 2}.
\ea
If the parameter $a=1$ then the action (\ref{dilaton}) corresponds
to the CGHS model \cite{CGHS}. A static solution describing the CGHS dilaton
black hole is
\ba\nonumber
ds^2= -zdt^2 +{dz^2\over z(1-z)^2}.
\ea
with the horizon located at $z=0$ and spatial infinity at $z=1$.

%%%%%%%%%%%%%%%%%%%%%%%%%%%%%%%%%%%%%%%%%

\section{Euclidean dilaton black hole}\label{sec_2}

\begin{figure}
\centerline{\includegraphics[width=7cm, height=4cm]{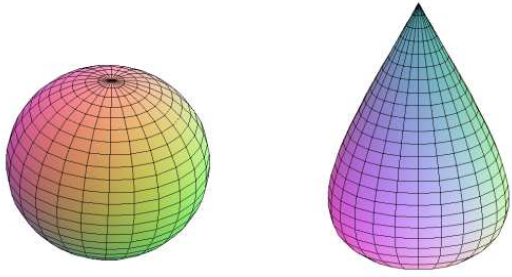}}
\caption{Embeddings of the Euclidean $a=2$ (on the left) and $a=3/2$
 (on the right) geometries to a flat 3D space.}
\label{fig3}
\end{figure}

\begin{figure}
\centerline{\includegraphics[width=7cm, height=4cm]{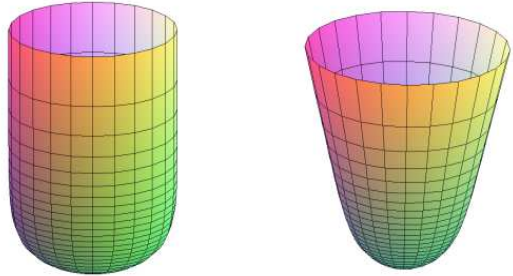}}
\caption{Embeddings of the Euclidean CGHS black hole (on the left, $a=1$)
and a dilaton black hole with $a=1/2$ (on the right) to a flat 3D space.}
\label{fig4}
\end{figure}

Let us start with the Euclidean version of the BH. It is obtained
from the metric (\ref{metric_1}) using
the Wick rotation of the Killing time $t=i\tau$, the Euclidean time
being periodic with the period $\beta=1/T_{\mathrm BH}=4\pi$.

\ba\n{Eucl_metric}
ds^2=z(1-z)^{a-1}d\tau^2+{(1-z)^{a-3}\over z}~dz^2, \hskip 1cm
0\le\tau\le4\pi   ,\hskip 0.5cm
0\le z\le 1.
\ea
In the Euclidean case it is useful to rewrite the metric in another
coordinate system $(\theta,\varphi)$
\ba\nonumber
\cos(\theta)=1-2z,\hskip 1cm
\varphi={1\over 2}\tau.
\ea
in which it is explicitly conformal to a sphere $S^2$.
\ba
ds^2=\left({1+\cos\left(\theta\right)\over 2}\right)^{a-2}
\left[d\theta^2+\sin(\theta)^2~d\varphi^2\right], \hskip 1cm
0\le \theta\le\pi, \hskip 0.5cm 0\le\varphi\le2\pi.
\ea
The corresponding scalar curvature for this geometry is
\ba\nonumber
R=a\left({1+\cos\theta\over 2}\right)^{2-a}
,\hskip 1cm
\sqrt{g}R=a\sin\theta
\ea
At the point $\theta=\pi$ we have to be more accurate. If $1<a<2$ and the point
$\theta=\pi$ is not excluded from the manifold then there is a conical singularity
at this point (see, e.g., the $a=3/2$ case right picture in figure \ref{fig3}) and
one has to add $\delta$-like singular curvature term
\footnote{For the Minkowskian signature $\delta$-like term does not appear and,
therefore, it is of no consequence for our consideration of Feynmann propagators.}
\ba\nonumber
R=\left({1+\cos\theta\over 2}\right)^{2-a}
\left[a+(2-a)\delta\left({1+\cos\theta\over 2}\right)\right].
\ea\\
For compact manifolds the inclusion of conical singularity term guarantees the
correct value of the
topological invariant $\int d^2x\sqrt{g}R=8\pi$. Embeddings of these manifolds
to a 3D flat space are depicted in figure (\ref{fig3}).
When $a=2$ the conical singularity disappears and we have the metric of a
unit sphere, describing, obviously, an analytic continuation of the de Sitter space.
If $0< a\le 1$ then the Euclidean manifolds have the topology of
a disk (see figure \ref{fig4}) and conical singularities do not appear.

%%%%%%%%%%%%%%%%%%%%%%%%%%%%%%%%%%%%%%%%%%%%%%%%%%%%%

\section{Euclidean vacuum}

Now let us consider quantum fields, $\phi$, which are non-minimally coupled to the
scalar curvature on this background.
The equation for free fields in 2D can always be written in the form
\ba\nonumber
\left(\Box-\xi R\right)\phi=0
\ea
The Euclidean Green function is the solution of the equation
\ba
\left(\Box-\xi R\right)G_E(\theta,\varphi;\theta',\varphi')
=-{\delta(\theta-\theta')\delta(\varphi-\varphi')\over\sqrt{g}}.
\ea
When quantum fields are assumed to be regular at ``south pole'' ($\theta=\pi$)
it's explicit form reads
\ba
\left[\partial_{\theta}\left(\sin\theta~\partial_{\theta}\right)
+{1\over\sin\theta}\partial^2_{\varphi}-a\xi\sin\theta\right]
G_E(\theta,\varphi;\theta',\varphi')
=-\delta(\theta-\theta')\delta(\varphi-\varphi').
\ea
A remarkable property of this Green function is that the parameter $a$ enters the
equation only in the combination $a\xi$.
If we put $a\xi=2\tilde{\xi}+m^2$ then this equation is identical to that of
a massive non-conformal scalar field on the 2D sphere
\ba\nonumber
\left(\Box-\tilde{\xi}
R-m^2\right)G_E(\theta,\varphi;\theta',\varphi')\Big|_{{\cal M}=S^2}
=-{\delta(\theta-\theta')\delta(\varphi-\varphi')\over\sqrt{g}}.
\ea
It means that in spite of the quite different geometries of the spacetimes in question non-minimal
quantum fields do not distinguish between them after the rescaling of the constant $a\xi$.

The Green function for the non-minimal scalar field in the background
of the Euclidean CGHS black hole $(a=1)$ has been found in
\cite{FroZel:2001} (see Eqs.(5.30)-(5.31)).
In order to make more transparent the comparison of our results with
the de Sitter case \cite{Polyakov:2007} and other dilaton black holes
we define $a\xi=-\nu(\nu+1)$ or
\ba\n{nu}
\nu=-{1\over 2}+\sqrt{{1\over 4}-a~\xi}.
\ea
Then, for arbitrary $a>0$ in the metric (\ref{metric_1}) the Euclidean  Green
function can be written in terms of the Legendre function exactly as for
CGHS black hole case (see \cite{FroZel:2001})
\ba\n{G_E}
G_E(X,X') = -{1\over 4\sin(\pi\nu)}~P_{\nu}(-\lambda),
\ea
where
\ba\nonumber
\lambda&=&(1-2z)(1-2z')+4\sqrt{zz'(1-z)(1-z')}\cos\left({\tau-\tau'\over
2}\right)\\ \n{lambda}
&=&\cos\theta\cos\theta'+\sin\theta\sin\theta'\cos(\varphi-\varphi').
\ea
The dependence on the geometry described by the constant $a$ comes only
via the parameter $\nu$.
$G_E(X,X')$ is regular at ``antipodal" points ($\lambda= -1$) and has a proper
logarithmic divergence at coincident points ($\lambda\rightarrow 1$).
Strictly speaking $\lambda= -1$ means that points are antipodal only
in the 2D sphere case. But the coordinate dependence of Green functions for all other
spaces is also encoded in the universal function $\lambda(X,X')$ which
in general is no longer a trivial function of the geodesic distance
between points.

The mode expansion of the Green function can be written in the form
\begin{eqnarray}
G_E(X,X') &=& -{1\over 4\sin(\pi \nu)}~\Big[~
P_{\nu}(-1+2 z^<)~P_{\nu}(1-2 z^>)  \nonumber \\
&+& 2\sum_{n=1}^\infty (-1)^n\cos\left({n(\tau-\tau')\over 2}\right)
~P^{~n}_{\nu}(-1+2 z^<)~P^{-n}_{\nu}(1-2 z^>)
\Big] ~.
\end{eqnarray}
After an analytical continuation to the Minkowskian signature this Green function
gives the Feynmann propagator for quantum fields in the Euclidean vacuum state.
The de Sitter Green function corresponds to $a=2$ and CGHS black hole case
to $a=1$. Of course in the Minkowskian signature one can consider a set of different
vacuum states and the Feynmann propagator depends on their choice. The Euclidean vacuum is
only one of the possibilities. Moreover, if one wants to
take into account interactions, there are additional constraints on a
possible choice of the vacuum state. In order to get a meaningful perturbation theory
the Feynmann propagator should satisfy the composition principle \cite{Polyakov:2007}.

%%%%%%%%%%%%%%%%%%%%%%%%%%%%%%%%%%%%%%%%%%%%%%%%%%%%%

\section{Composition principle}

The variations of both the Euclidean Green function $G_E$ and
the Feynmann propagator should satisfy the following rule
\ba\nonumber
\delta G = G\delta\hat{F}G,
\ea
where
\ba\nonumber
\hat{F}G=-\hat{1},\hskip
1cm\Leftrightarrow
\hskip 1cm
\sqrt{g}\hat{F}G(x,x')=-\delta(x-x').
\ea
So, if we consider the following variation of the operator
$\delta\hat{F}=-R\delta\xi$,
then the Green functions must fulfill the identity
\ba\n{composition_1}
{\partial}_{\xi}G(x,y) = -\int dx'\sqrt{g(x')}~G(x,x')R(x')G(x',y)
\ea
In the case of a 2D unit sphere, which corresponds to $a=2$ and
the scalar curvature $R=2$, this variational rule
is equivalent to the composition principle
\ba\n{composition_2}
{1\over 2}{\partial}_{\tilde{\xi}}G(x,y)=
{\partial}_{m^2}G(x,y)
= -\int dx'\sqrt{g(x')}~G(x,x')G(x',y)
\ea
proposed by Polyakov as a property of Green functions
necessary for the quantum state to be ``eternal" \cite{Polyakov:2007}. The
composition principle lies at the foundation of quantum field theory
and its violation would be unacceptable for any quantum field theory
which takes into account interactions.

In the case of a 2D Euclidean dilatonic black hole the Ricci scalar is
not constant, but $\xi
R\sqrt{g}\big|_{EBH}\leftrightarrow({2\tilde{\xi}+m^2})
\sqrt{g}\big|_{S^2}$. Therefore, taking into account the coincidence
of equations for Green functions in both spaces we see that the
``eternity" condition (\ref{composition_2}) formally coincides with
\eq{composition_1}. For the Euclidean Green functions  \eq{G_E}
this property \eq{composition_1} naturally follows from the property
of the Legendre functions.

The Feynmann propagator can be obtained formally
by analytic continuation from the Euclidean Green function.
\ba\n{G_F}
G_{E}(X,X')=-{1\over 4\sin(\pi\nu)}~P_{\nu}(-\lambda-i0)
\ea
This propagator corresponds to a particular choice of the
vacuum state. In the case of the de Sitter spacetime it is known as the
Bunch-Davies or Hartle-Hawking vacuum. Sometimes this state is also
called the Euclidean vacuum.
This is why we have kept here the subscript $E$ though it is defined for
spacetimes with the Minkowskian signature.

Let us consider an example of 2D de Sitter spacetime.
In global coordinates it has the form
\ba\nonumber
ds^2=-d\tau^2+\cosh^2{\tau}d\tilde{\varphi}^2,\hskip 1cm
-\infty<\tau<\infty,\hskip 1cm 0\le\tilde{\varphi}\le2\pi,
\ea
and the de Sitter invariant quantity
\ba\nonumber
\lambda=\cosh\tau\cosh\tau'\cos(\tilde{\varphi}-\tilde{\varphi}')-
\sinh\tau\sinh\tau',
\ea
is greater than $1$ for timelike separations and less then $1$ for
spacelike separations.
As a result of the different range of integration in the
\eq{composition_2} the composition rule is violated for
the Euclidean vacuum \cite{Polyakov:2007}
since the Legendre function $P_{\nu}(-\lambda)$
blows up in the asymptotic $\lambda\rightarrow\infty$  and the
integral over the spacetime diverges.
The Euclidean vacuum is not the only de Sitter
invariant vacuum. The Green functions for other de Sitter
invariant vacuum states
can be written as a linear combination of $P_{\nu}(\lambda+i0)$ and
$Q_{\nu}(\lambda+i0)$. Here $i0$ is added to the argument
of the Legendre functions in
order to define the propagator on the cut  $[-\infty, 1]$ in the
complex plane of $\lambda$.
Polyakov proposed \cite{Polyakov:2007} that the Feynmann
propagator which satisfies the composition principle for the de Sitter
space is given by the formula
\ba\n{G_Q}
G_{Q}(X,X') &=& {1\over2\pi}Q_{\nu}(\lambda+i0).
\ea
It corresponds to a different vacuum state and
differs from the Euclidean vacuum propagator. The difference can be
written explicitly
\ba\n{G_Q1}
G_{Q}(X,X')&=&G_E(X,X')+{{\mathrm e}^{-i\pi\nu}\over 4\sin(\pi\nu)}~P_{\nu}(\lambda+i0)
\ea
The Green function \eq{G_Q} decreases  at $\lambda\rightarrow\infty$
and the integral in  \eq{composition_2} converges provided
${\Re}(\nu)>-1/2$.
$G_{Q}$ has a proper divergence at coincident points and there is also an extra
divergence for antipodal points. Evidently, the
corresponding  quantum state belongs to the class of $\alpha-$vacua
\cite{Mottola:1985,Collins:2003,Collins:2004,Kaloper:2002,EinhornLarsen:2003,BanksMannelli:2003}.
In the de Sitter spacetime $G_{Q}$ is the only Feynmann propagator which respects the
composition rule. Nevertheless, even this vacuum state is not stable and decays.
This instability can be proved \cite{Polyakov:2007} using analytic properties
of $Q_{\nu}(\lambda)$.

Though the generic spacetimes \eq{metric} do not respect the de Sitter symmetry,
the Green functions still have the form \eq{G_Q} of the Feynmann propagators
in the de Sitter
spacetime. Therefore, the above discussion of their properties is applicable to all
these spacetimes and one can conclude that Q-vacuum for dilaton black holes
decays as well.

%%%%%%%%%%%%%%%%%%%%%%%%%%%%%%%%%%%%%%%%%%%%%

\section{Bound states, quasinormal modes and spacetime instability}\label{bound}

In static coordinates $(t,r)$ (see \eq{metric_1})
Fourier modes $\phi_{\omega}$ with fixed frequency $\omega$ satisfy the equation
\ba
\left[\partial_r(1-r^2)\partial_r +\frac{4\omega^2}{1-r^2}-a\xi\right]\phi_{\omega}=0
\ea
The independent solutions of this equation are
$P^{2i\omega}_\nu(r)$ and $Q^{2i\omega}_\nu(r)$.
The Euclidean propagator $G_E$ is the sum of terms
\ba\nonumber
{\mathrm e}^{-i\omega(t-t')}P^{2i\omega}_\nu(r)~P^{-2i\omega}_\nu(-r')
\ea
with a prefactors proportional to the thermal occupation number
$n(\omega)={1\over {\mathrm e}^{4\pi\omega}-1}$
corresponding to the inverse temperature $\beta=2\pi/\kappa=4\pi$.
The $G_Q$ propagator is a similar sum of terms
\ba\nonumber
{\mathrm e}^{-i\omega(t-t')}P^{2i\omega}_\nu(r)~Q^{-2i\omega}_\nu(r')
\ea

Let us study an evolution of the modes with various boundary conditions.
Any mode with a frequency $\omega$ is given by a linear
combination of the Legendre functions $P^{2i\omega}_\nu(r)$ and
$Q^{2i\omega}_\nu(r)$ or they can be equally well expressed in terms of
hypergeometric functions (see, e.g., \cite{FroZel:2001}).
The frequency in these formulas can be taken as any complex number
depending on the boundary conditions imposed on the modes.
The solutions with real frequencies describe wavelike excitations.
Complex frequencies are related to quasinormal modes while
modes with pure imaginary frequencies appear to describe bound states.
In the paper \cite{FroZel:2001} it has been shown that for a CGHS
black hole bound states appear when $\xi$ is negative.
In this case perturbations grow exponentially with time and lead to a
severe "tachionic" instability of the black hole. It's quite interesting that
the unstable modes appear in the region outside the horizon, where the
potential barrier has a minimum. Bound modes create fluxes of energy from
this region to infinity and to the horizon and, because of the back reaction,
eventually strongly deform the background geometry.

In our case the modes are determined by the same
formula (4.7) from the paper \cite{FroZel:2001}, but $a\xi$ being
substituted for $\xi$.
\ba\n{omega_n}
\omega_n={i\over 2}\left(\nu-n\right)
={i\over 2}\left(\sqrt{{1\over 4}-a\xi}-{1\over 2}-n\right)
\ea
When $a\xi$ is negative and decreases further the number of bound states
increases. A new bound state appears every time when
$\nu=\sqrt{{1\over 4}-a\xi}-{1\over 2}$ reaches a new integer number value.
This automatically leads to the conclusion that quantum fields with
negative $a\xi$ cause "tachionic" instability for all spacetimes described
by the metrics \eq{metric} including the dilaton black holes and the de Sitter space.

The modes which are ingoing at the horizon $\sim{\mathrm e}^{-i\omega(t+x^*)}$
and outgoing at infinity $\sim{\mathrm e}^{-i\omega(t-x^*)}$ are
called quasinormal modes (QNM). The parameters of QNMs can be easily
determined from the position of poles in the complex $\omega$-plane
of a transmission coefficient (see, e.g., Eq.(3.30) in \cite{FroZel:2001})
\ba
|T_{\omega}|^2={\cosh(4\pi\omega)-1\over
\cosh(4\pi\omega)+\cos\left(2\pi\sqrt{{1\over 4}-a\xi}\right)}
\ea
through the potential barrier.
When the frequency is pure imaginary
with positive imaginary part they describe bound states
\eq{omega_n}. The unstable behavior of
2D black holes against scalar perturbations discussed
in \cite{Becar:2007} is directly related with the instability
because of the bound states \cite{FroZel:2001}.

For other values of $a\xi$ the frequencies \eq{omega_n} may have both
imaginary and real parts. If $a\xi>{1\over 4}$, then the
quasinormal modes $\omega_n$ are
\ba\n{QNM}
\omega_n={1\over 2}\sqrt{a\xi-{1\over 4}}-i~{1+2n\over 4}.
\ea
Here we see that the real part
$\Re(\omega_n)={1\over 2}\sqrt{a\xi-{1\over 4}}$ of the quasinormal modes
does not depend on $n$ in accord with the high damping limit conjecture
\cite{Daghigh:2005,Hod:1998} for QNMs.
In pure de Sitter spacetime ($a=2,~2\xi=m^2$) quasinormal modes \eq{omega_n}
correspond to those of 2D case in \cite{Du:2004,Lopes-Ortega:2006}.

%%%%%%%%%%%%%%%%%%%%%%%%%%%%%%%%%%%%%%%%%%%%%%%%%%%%%%%%%%%%%%%

\section{Discussion}\label{discussion}

2D dilaton gravity models appear as effective gravity
theories after a dimensional reduction from higher dimensions, in string theory,
and in many other applications. The advantage of studying quantum fields in 2D
spacetimes is that it is much easier to find exact solutions in this case, while
reproducing qualitatively the same physical effects such as the Hawking radiation,
thermodynamics etc.. Well known examples are CGHS model \cite{CGHS} and the de
Sitter spacetime. In this paper we have found that these two models are related.
Moreover they, in fact, have many more relatives. As for the non-conformal quantum
fields on these backgrounds, they are exactly solvable for all members of this family
and their physical properties are basically the same.
The difference is encoded only in one real parameter $a$ \eq{metric}, which marks the
representative of the family.

For example, exact quasinormal modes are described by the common formulae
Eqs.(\ref{omega_n}), (\ref{QNM}). It is surprising because the geometries of these
spacetimes are quite different. From this result one can see that there is no
real part for quasinormal modes as soon as $\xi\le{1\over 4a}$. When
$\xi\ge{1\over 4a}$ a real part appears and it is the same for all $n$.
The real part ${1\over 2}\sqrt{a\xi-{1\over 4}}$ can be any real number, while
the Hawking temperature is the same for all these spacetimes.
This observation can be considered as a counterexample to the conjecture, that
the real part of highly damped QNMs is to be proportional to the logarithm
of an integer number. This conjecture has been widely discussed in the literature
in relation with the black holes quantization (see, e.g., review of the problem and
references on the subject in the paper by R.G. Daghigh and G. Kunstatter
\cite{Daghigh:2005}).

If scalar fields have negative coupling $\xi$ then bound modes arise outside
the horizon of CGHS black hole \cite{FroZel:2001}. This lead to classical
instability of the system similar to tachionic one. The observed equivalence
of CGHS black hole, de Sitter spacetime and other dilaton models leads to the
conclusion that as soon as $a\xi<0$ bound modes appear for all these metrics
along with the same destructive instabilities. For the positive non-minimal coupling
quantum instabilities appear to arise similar to instabilities discussed by Polyakov
\cite{Polyakov:2007} in application to the de Sitter. Their interpretation
in the black hole case requires more detailed analysis and would be very
interesting to examine.

%%%%%%%%%%%%%%%%%%%%%%%%%%%%%%%%%%%%%%%%%%%%%%%%%%%%%%%%%%%%%%%

\acknowledgments  It is a pleasure to thank Valeri Frolov for
inspiration and useful discussions. This work was  partly supported
by  the Natural Sciences and Engineering Research Council of Canada. The
author is grateful to the Killam Trust for its financial support.

%%%%%%%%%%%%%%%%%%%%%%%%%%%%%%%%%%%%%%%%%%%%%%%%%%%%%%%%%%%%%%%

\end{document}